\documentclass{elsart1p}
\usepackage{graphicx}
\def\lsim{\mathrel{\rlap{\lower4pt\hbox{\hskip1pt$\sim$}}
    \raise1pt\hbox{$<$}}}         
\def\gsim{\mathrel{\rlap{\lower4pt\hbox{\hskip1pt$\sim$}}
    \raise1pt\hbox{$>$}}}         
\usepackage{amssymb}
\begin{document}
\begin{frontmatter}
%
%
%
\title{Fundamental Symmetries and Conservation Laws}
%
%
\author{W. C. Haxton}
\address{Inst. for Nuclear Theory and Dept. of Physics, 
  University of Washington, Seattle, WA 98195 USA}
\begin{abstract}
I discuss recent progress in low-energy tests of symmetries and conservation laws,
including parity nonconservation in atoms and nuclei, electric dipole moment
tests of time-reversal invariance, $\beta$-decay correlation studies, and decays
violating separate (family) and total lepton number.
\end{abstract}
\end{frontmatter}
%
\section{Parity Nonconservation (PNC)}
The use of parity in atomic spectroscopy dates from the 1920s, when it was 
introduced as a wave function label, prompting Wigner to demonstrate that such labeling
is a consequence of the mirror symmetry of the electromagnetic interaction.   Today parity 
and its violation are tools for probing aspects of the standard model (SM) (e.g., to
isolate the strangeness-conserving hadronic weak interaction) and
new physics beyond the SM (e.g., the contributions of a new boson $Z_0^\prime$ to
the running of weak couplings).

The weak interaction between atomic electrons and the nucleus is dominated by the coherent $A(e)-V(N)$
contribution.  As the
SM tree-level coupling to
protons, $c_V(p) = 1 - 4\sin^2 \theta_W \sim 0.1$, is suppressed while $c_V(n) = -1$,
the weak charge of the nucleus $Q_{weak} \sim Z c_V(p)+ N c_V(n)$ is
approximately $-N$.  Consequently 
\begin{equation}
H^{\mathrm{Atomic~PNC}}_{A(e)-V(N)} \sim {G_F \over 2 \sqrt{2}}~ Q_{weak}~ \rho_N(\vec{r}) ~\gamma_5,
\end{equation}
where $\rho_N(\vec{r})$ is the neutron density.  The effects of this short-range
interaction grow $\sim~Z^3$ and thus are most easily detected in heavy atoms.

Atomic PNC was first observed in 1978, while the best current limit comes from the 1997
JILA experiment,
$Q_{weak}$($^{133}$Cs) = $-73.16 \pm 0.29 \mathrm{(exp)} \pm 0.20 \mathrm{(theor)}$ \cite{JILA,
Derevianko}.  The $\sim 0.3$\%
precision poses a challenge for theoreticians attempting to calculate the
associated atomic mixing.  Advances in the atomic
structure calculations include improved evaluations of relativistic (Breit) and radiative corrections,
vacuum polarization, the neutron distribution, strong-field self-energies, and weak vertex corrections.
The corrections have been at the sub-1\% level for Cs, and in
total have brought SM calculations into agreement with experiment at $\lsim 1\sigma$.
Consequently beyond-the-SM contributions are constrained, yielding, e.g., a bound on the mass 
of an extra neutral boson, $M(Z_0^\prime) \gsim 1.3$ TeV \cite{Derevianko}.

\begin{figure}
\centering
\includegraphics[width=13.7cm]{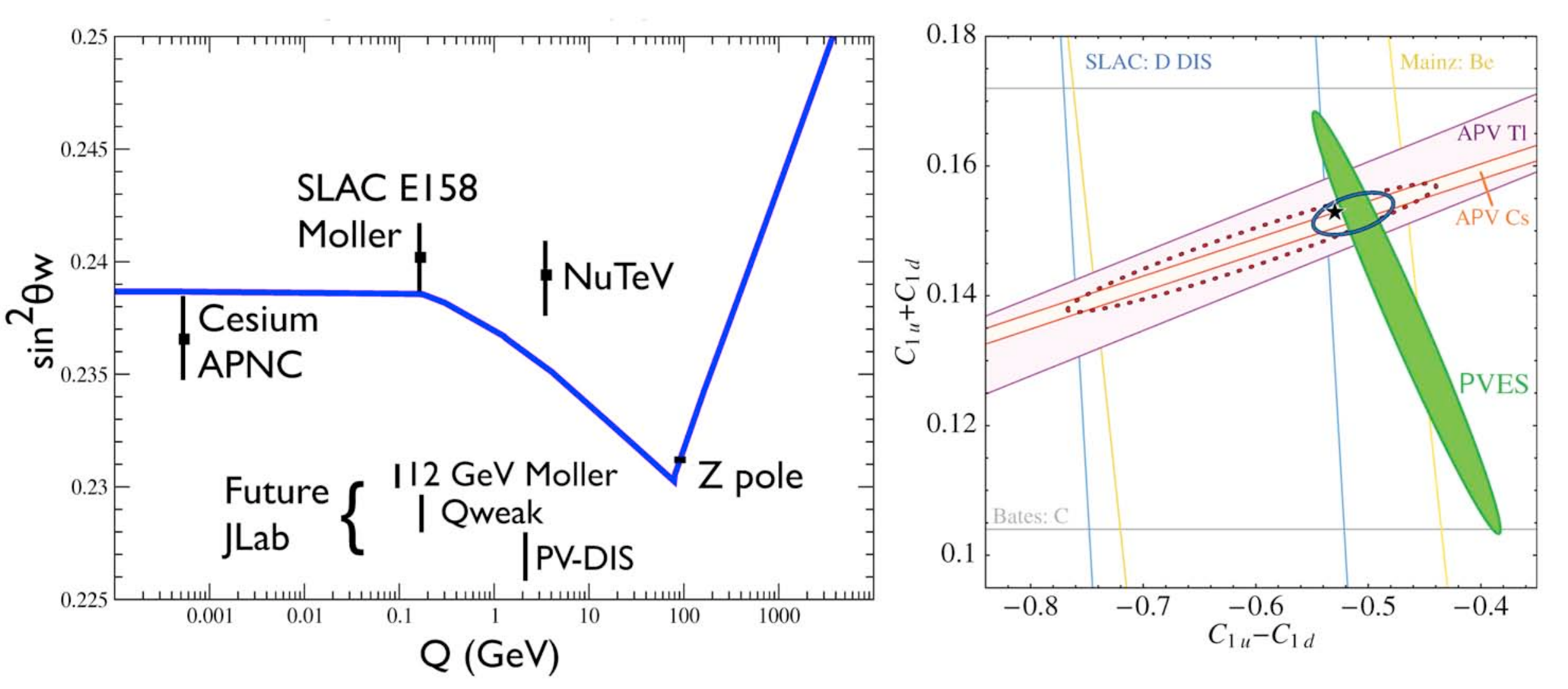}
\caption{In the left frame \cite{Erler} experimental determinations
of $\sin^2 \theta_W$ are compared to SM predictions, normalized
to the $Z$ pole.  The right frame \cite{Young} shows the
complementary constraints from atomic PNC ($Q_{weak} \sim N$)
and PNC electron scattering (where protons have been the favorite
target).}
\label{fig:PVES}
\end{figure}

The atomic PNC constraint on $c_V(p)$ and $c_V(n)$, or equivalently on the
underlying quark couplings, is essentially orthogonal to that from
PNC electron scattering experiments, as shown in the right frame of
Fig. \ref{fig:PVES} \cite{Young}.  In the left frame
experimental values for $\sin^2 \theta_W$ are superimposed on
SM predictions for its running \cite{Erler}.   There is good agreement.  Included on
this graph are the error bars experimentalist expect to achieve in future intermediate-energy
measurements at JLab.

\begin{figure}
\centering
\includegraphics[width=13.7cm]{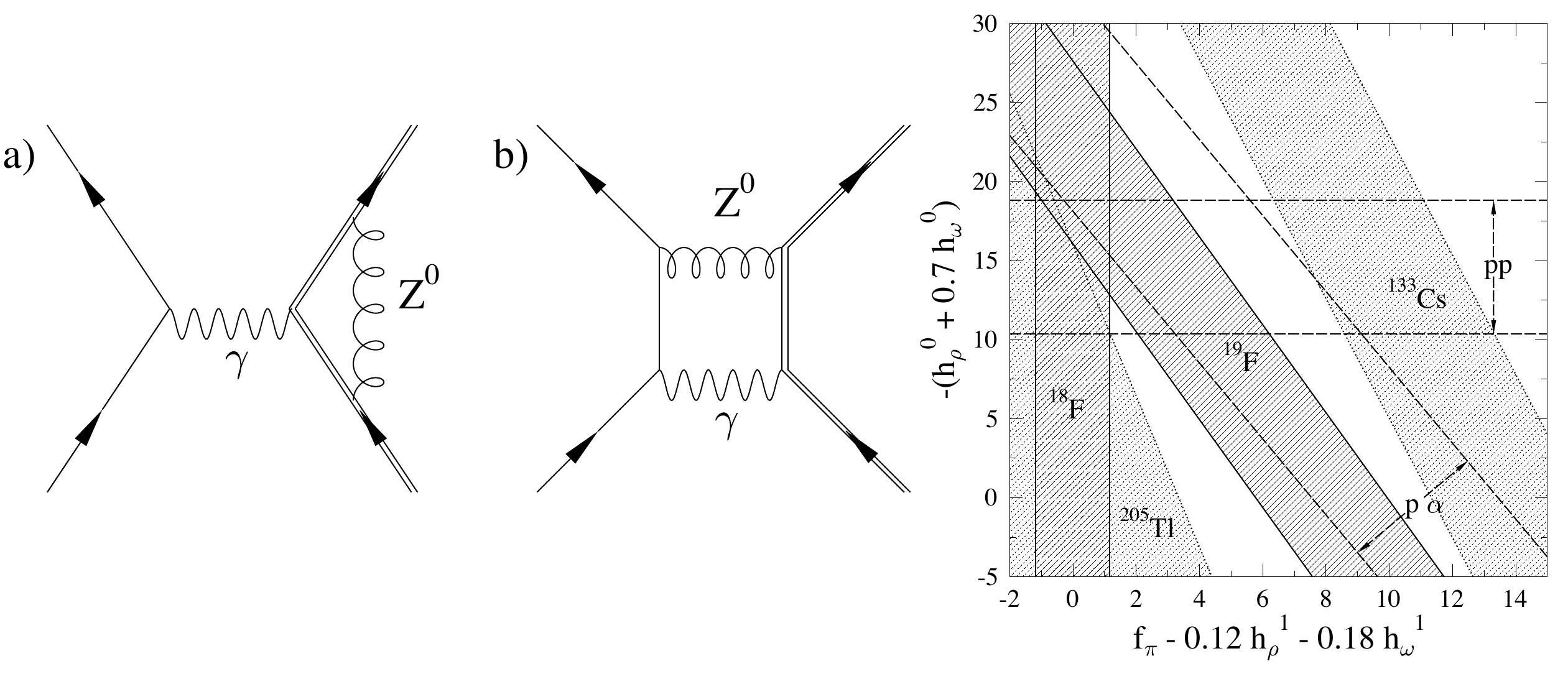}
\caption{The diagram on the left, a contribution to the anapole moment (a weak radiative
correction arising from single-photon exchange), is part of a larger class of such corrections,
including the box diagram at the center.  The anapole constraints from atomic PNC
studies of Cs and Tl are included in the diagram on the right, which summarizes the current status
of hadronic PNC studies.}
\label{fig:anapole}
\end{figure}

The Cs experimenters also studied the hyperfine dependence of the signal, from
which a small nuclear-spin-dependent contribution to PNC, $V(e)-A(N)$, 
was extracted,
\begin{equation}
H^{\mathrm{Atomic~PNC}}_{V(e)-A(N)} = {G_F \over \sqrt{2}} \kappa \vec{\alpha} \cdot \vec{I}\rho(\vec{r}) ~\Rightarrow~\kappa = \kappa_{Z_0} + \kappa_{\mathrm{HF}} + \kappa_{\mathrm{A}} = 0.112 \pm 0.016.
\end{equation}
The measured $\kappa$, obtained from 7000 hours of data,  constrains the sum of
three terms, tree-level $Z_0$ exchange
($\kappa_{Z_0} \sim 0.014$, suppressed because the nuclear coupling is no longer
coherent and the vector $Z_0$ coupling to the electron $\sim (4 \sin^2 \theta_W-1)/2 \sim -0.05$
is small), 
spin-dependent effects arising from the combination of hyperfine and $Q_{weak}$ interactions 
($\kappa_{\mathrm{HF}} \sim 0.0078$), and the nuclear anapole moment
($\kappa_{\mathrm{A}} \sim 0.090 \pm 0.016$), which
dominates the signal.  The nuclear anapole moment is a PNC coupling of a photon to
the nucleus (see Fig. \ref{fig:anapole}), part of a set of weak
radiative corrections.  It arises from a PNC torroidal current winding
within the nucleus, growing with atomic number as A$^{2/3}$ (proportional to the torroid's cross section).  This growth leads to a the weak
radiative correction that exceeds the tree-level axial $Z_0$ exchange in heavy nuclei.

Because the largest contribution to $\kappa_{\mathrm{A}}$ for a heavy nucleus comes from 
opposite-parity admixtures in the ground-state nuclear wave function, $\kappa_{\mathrm{A}}$
provides a constraint on hadronic PNC.  As discussed in the next section, the Cs
anapole moment seems to be somewhat larger than one would expect based on
other tests of hadronic PNC.

\section{The Nucleon-Nucleon Parity-Nonconserving Interaction}
\begin{figure}
\centering
\includegraphics[width=12cm]{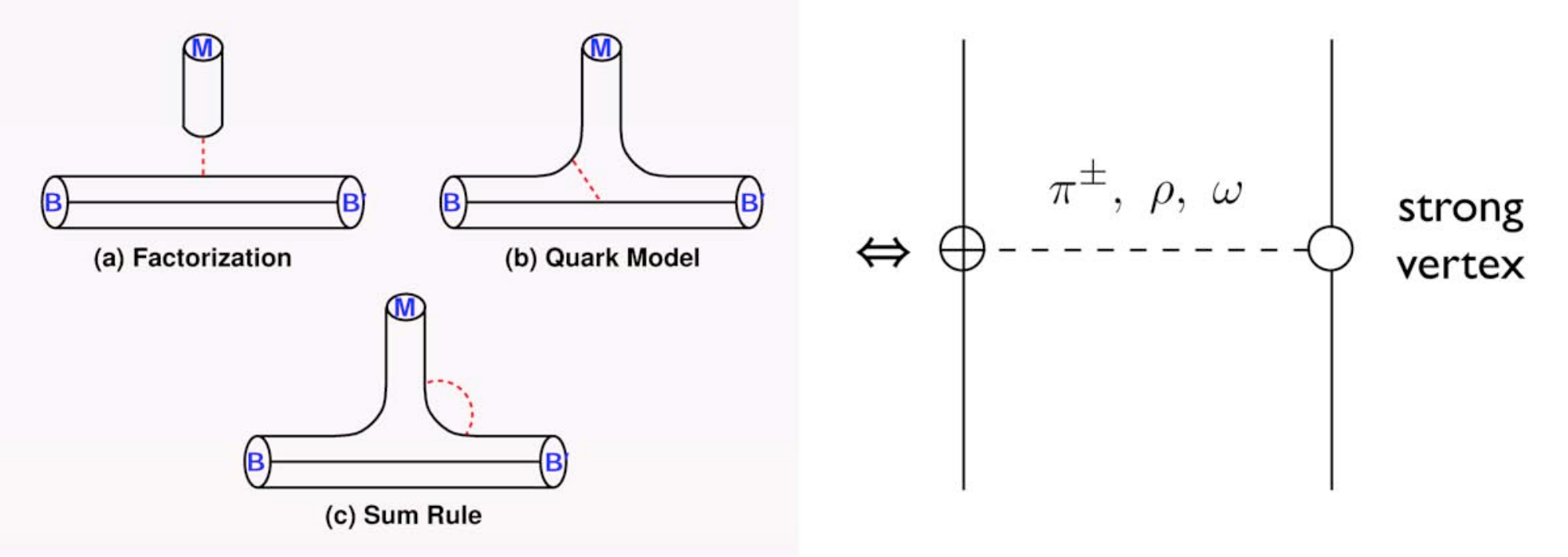}
\caption{A single-boson-exchange contribution to $V_{PNC}$ contains one weak vertex (left) and one strong one (right).  DDH \cite{DDH} related the weak vertex to SM quark currents, 
using factorization, the quark model, and sum rules.}
\label{fig:meson}
\end{figure}

The NN PNC interaction at low energies is characterized by the five S-P Danilov 
amplitudes of Table \ref{tab:PNC}, which in turn are often parameterized in terms of a
potential derived from $\rho$, $\omega$, and $\pi^{\pm}$ exchange \cite{DDH}.
This parameterization can be viewed as a phenomenological effective theory,
with the heavy meson exchanges playing the role of short-range interactions
$\vec{\nabla}_{12} \delta(\vec{r}_{12})$ in each of the five S-P channels, and with the pion
separately determining the long-range behavior of the potential
(see Fig. \ref{fig:meson}).  In fact, systematic
effective field theory formulations exist for
pionless theory and with explicit pions \cite{Dan}.

\begin{table}
\caption{S-P weak PNC amplitudes and the corresponding meson-exchanges \cite{Adelberger}}
\begin{tabular}{|l|c|c|c|c|c|c|}
\hline
~~~Transition~~~ & ~~~I $\leftrightarrow$ I$^\prime$ ~~~&~~~ $\Delta$I ~~~&~~~ n-n ~~~&~~~ n-p ~~~&~~~ p-p~~~ &~~~ meson exchanges ~~~ \\
\hline
${}^3S_1 \leftrightarrow {}^1P_1$ & 0 $\leftrightarrow$ 0 & 0 & & x & & $\rho,\omega$ \\
${}^1S_0 \leftrightarrow {}^3P_0$ & 1 $\leftrightarrow$ 1 & 0 & x & x & x & $\rho,\omega$ \\
 & & 1 & x & & x & $\rho,\omega$\\
 & & 2 & x & x & x & $\rho$ \\
 ${}^3S_1 \leftrightarrow {}^3P_1$ & 0 $\leftrightarrow$ 1 & 1 & & x & & $\pi^\pm,\rho,\omega$ \\
 \hline
\end{tabular}

\label{tab:PNC}
\end{table}

One of the goals of the field has been to isolate the neutral current contribution to
hadronic PNC.  While the weak interaction can be observed in
flavor-changing hadronic decays, the neutral current contribution to such
decays is suppressed by the GIM mechanism and thus unobservable.
The NN and nuclear systems are thus the only practical laboratories for studying
the hadronic weak interaction in all of its aspects \cite{Adelberger}.

As the weak contribution to the NN interaction is much
smaller than the strong and electromagnetic contributions,
PNC is  exploited to isolate weak effects.   The most common
observables are pseudoscalars arising from the interference of weak and
strong amplitudes, e.g., the circular polarization of $\gamma$ rays emitted
from an unpolarized excited nuclear state, or the $\gamma$ ray asymmetry 
if the nuclear state can be polarized.  As the observable depends on a product
of parity-conserving and PNC amplitudes,
the weak interaction appears linearly.  Alternatively, there are processes, such as
the $\alpha$ decay of an unnatural-parity state to a 0$^+$ final
state, where the observable depends on the square
of a weak amplitude, and consequently is not a pseudoscalar.

The isospin of meson-nucleon couplings has an interesting relation to the underlying bare
charged and neutral currents. The hadronic weak interaction is
\begin{equation}
L^{eff} = {G \over \sqrt{2}} \left[ J^\dagger_W J_W + J_Z^\dagger J_Z \right] + h.c.~~~~~~~~
J_W = \cos{\theta_C} J_W^{\Delta S=0} + \sin{\theta_C} J_W^{\Delta S =-1},
\end{equation}
where the charge-changing current is the sum of $\Delta$I=1 $\Delta$S=0 and
$\Delta$I=1/2 $\Delta$S=-1 terms.
Consequently the $\Delta$S=0 interaction has the form
\begin{equation}
L^{eff}_{\Delta S=0} = {G \over \sqrt{2}} \left[ \cos^2{\theta_C}J_W^{0 \dagger} J_W^0 +
\sin^2{\theta_C} J_W^{1 \dagger} J_W^1+ J_Z^\dagger J_Z \right]
\end{equation}
where the first term, a symmetric product of $\Delta$I=1 currents, has $\Delta$I=0,2, while the
second term, a symmetric product of $\Delta$I=1/2 currents, is $\Delta$I=1 but
Cabibbo suppressed.  Consequently a $\Delta$I=1 PNC meson-nucleon vertex should
be dominated by the neutral current term -- a term not accessible in strangeness-changing
processes.  This is the $\pi^\pm$ exchange channel.
One could isolate this term by an isospin analysis of a complete
set of PNC NN observables or, alternatively, by finding a case in which the isospins of the
admixed nuclear states select only the $\Delta$I=1 contribution

Unfortunately, experimental progress in this field has been slow.  Ideally one would like to
avoid constraints from complex nuclei, as calculated nuclear mixing matrix elements 
are generally needed in the analysis.   But there too few constraints available from
$NN$ and few-body experiments.  The latest effort to improve this situation, the LANSCE
measurement of the analyzing power $A_\gamma(\vec{n}+ p \rightarrow d + \gamma)$,
fell short of its goal and thus will need to be revisited when more intense neutron
sources become available at the SNS.  The best available data are \cite{Adelberger}
\begin{flushleft}
\[A_L(\vec{p}+p, 45~\mathrm{MeV}) = (-1.57 \pm 0.23) \times 10^{-7}~~~~~P_\gamma(^{18}\mathrm{F}) = (12 \pm 38) \times 10^{-5} ~~~ \]
\[A_L(\vec{p}+\alpha, 46~\mathrm{MeV}) = (-3.34 \pm 0.93) \times 10^{-7} ~~~~~A_\gamma(^{19}\mathrm{F} )= (-7.4 \pm 1.9) \times 10^{-5} \]
\end{flushleft}
The mixing of nearly degenerate opposite parity doublets accounts for the large
nuclear PNC signals.
$P_\gamma(^{18}$F) is important because 1) the mixing is isovector and
2) the nuclear matrix element can be extracted from ancillary measurements
(related axial-charge $\beta$ decay) with very little uncertainty.  The $^{19}$F case
is also partly constrained by similar data.  Finally, there is the Cs anapole
moment where, despite the apparent complexity of the nuclear structure physics, the
existing theoretical analyses are in good accord.

The analyses of these data can be displayed as constraints on effective isoscalar and
isovector weak couplings, as shown
in the right panel of Fig. \ref{fig:anapole}.  If the Cs anapole moment is
excluded, there is a region of overlap, corresponding
to a small isovector coupling (compared to the ``best value" of \cite{DDH}) and an
isoscalar coupling $\gsim$ the DDH ``best value."  The
isovector coupling is constrained by the $^{18}$F measurement and consistent with
zero: this component is the test for neutral currents.   But the
conclusion of a suppressed isovector NN PNC interaction rests entire on this one 
measurement and associated analysis.  When the $^{133}$Cs anapole moment is added, 
no solution is found: the anapole moment appears to be larger than one would have
expected, based on direct measurements of hadronic PNC.  A small corner of
the upper bound on the anapole moment of $^{205}$Tl is also shown:  while the
error bar on the Tl measurement is quite large, the result favors a coupling opposite
to that of $^{133}$Cs, contradicting theory expectations.

Clearly the field needs a new generation of higher precision experiments, including
neutron observables (e.g., in the $n+p$ system or $\vec{n}$ + $^4$He), to
make progress.

\section{Electric Dipole Moments and CP Violation}
A permanent electric dipole moment $d$ of an elementary particle or of a composite system
(such as an atom) requires both time-reversal and parity violation: 
$H_{edm} = d~ \vec{E} \cdot \vec{s}$ reverses sign under $t \rightarrow -t$ and under
$\vec{r} \rightarrow -\vec{r}$.   The signature of an edm is precession of the particle's
spin around the direction of the applied field, with a frequency proportional to $d$
and to the strength of the applied electric field.  By the CPT theorem, a nonzero
T-violating edm implies CP nonconservation (CPNC). Searches for edms
\begin{itemize}
\item test the SM's two sources of CPNC, the CKM phase
measured in neutral kaon decays and the unmeasured QCD $\theta$ parameter 
(edm searches require $| \theta | \lsim 10^{-10}$); and
\item probe CPNC beyond the SM (baryogenesis appears to
require new sources).
\end{itemize}
The sensitivity of measurements is remarkable.  The dipole moment limit for $^{199}$Hg
corresponds to a strain of about 10$^{-19}$, if interpreted in terms of a linear
displacement.  That is, were one to expand the atom to
the dimensions of the earth, such an edm corresponds to a displacement of
initially overlapping  uniformly charged spheres (+ and -)
by 10$^{-4}$ angstroms.  The Hg precession sensitivity, given typical electric
fields of $\sim 10^5$ v/m, corresponds to shifts in energy level splittings of $\sim 10^{-26}$ eV.

The connection between experimental limits and fundamental Lagrangians is
generally not simple.  Particularly in the case of the edms of diamagnetic atoms --
where the spin is carried by the nucleus -- a theorist must relate the fundamental CPNC phases
to effective low-energy couplings, determine the nuclear interactions these couplings
induce, calculate the resulting CPNC mixing of nuclear states and thus the
nuclear edm, and finally evaluate the atomic screening effects that determine the
residual atomic edm, the quantity experimentalists measure.

The general electromagnetic current for a spin-1/2 fermion $\langle p^\prime | j_\mu^{em} | p \rangle$  is
\begin{equation}
 \bar{N}(p^\prime) \left( F_1(q^2) \gamma_\mu
+ F_2(q^2) \sigma_{\mu \nu} q^\nu + {a(q^2) \over m^2} (\not q q_\mu - q^2 \gamma_\mu) \gamma_5
+ d(q^2) \sigma_{\mu \nu} q^\nu \gamma_5 \right) N(p),
\end{equation}
where $F_1$ and $F_2$ are the ordinary  charge and magnetic couplings, $a$ is the
anapole coupling, and $d$ the edm.  In an atom or nucleus, this last term will generate
odd static charge multipoles (C1 (the edm), C3, ...) and even static magnetic multipoles (M2, ...)
that are CPNC and PNC (provide the spin allows M2 and C3 moments).

Edm experiments have been done on various neutral systems,
including free neutrons, paramagnetic atoms or molecules
(the unpaired electron provides sensitivity to the electron edm), and diamagnetic atoms (paired
electrons but a nonzero nuclear spin, so that the valence nucleon's edm
and CPNC nuclear state mixing can be probed).  Table \ref{tab:edm} lists some of the
bounds.  The field is quite active, with new or proposed efforts
including ultracold neutrons
(Ill, PSI, Munich, SNS), $^{199}$Hg (Seattle), liquid $^{129}$Xe (Princeton), trapped
$^{225}$Ra (Argonne, KVI) and $^{213}$Ra (KVI), trapped $^{223}$Rn, and the deuteron
(BNL ring experiment).   Future cold-neutron efforts, for example, may improve
the current upper bound, 2.9 $\times 10^{-26}$ e cm, by a factor $\sim$ 60 in the
next decade \cite{CERN}.

\begin{table}
\caption{Edm limits, direct or derived, and expected SM level, based on the CKM phase \cite{CERN,Hg}.}
\begin{tabular}{|l|c|c|c|}
\hline
~Particle~ & ~edm limit~ & ~System~ & ~SM prediction (CKM phase)~ \\
\hline
~e & ~1.9$ \times 10^{-27}$ e cm~~& ~atomic $^{205}$Tl ~& ~$10^{-38}$ e cm~\\
~p & ~6.5$ \times 10^{-23}$ e cm~~& ~molecular TlF~ &~$10^{-31}$ e cm~\\
~n & ~2.9$ \times 10^{-26}$ e cm~~& ~ultracold n~ &~$10^{-31}$ e cm~\\
~$^{199}$Hg & ~3.1$ \times 10^{-29}$ e cm~~& ~atom vapor cell~ &~$10^{-33}$ e cm~\\
\hline
\end{tabular}
\label{tab:edm}
\end{table}

A new result for  $^{199}$Hg, anticipated when this talk was delivered, has been announced.
The experiment uses $\sim 10^{14}$ neutral atoms in a vapor cell
designed to extend the spin relaxation time for Hg
(which has atomic spin 0) to 100-200 s.  The edm resides on the nucleus, which is shielded
from an applied field by the polarization of the atomic cloud.  Consequently a net interaction
energy is generated only through nuclear finite-size effects, resulting in a reduced sensitivity
to the nuclear edm,
\begin{equation}
d_{\mathrm{atomic}} \sim 10 Z^2 \left( {R_N \over R_A} \right)^2 d_{\mathrm{nucleus}},
\end{equation}
where $R_N$ and $R_A$ are the nuclear and atomic sizes.  Such Schiff-shielding effects are less severe
in heavy atoms, because of the stronger Coulomb field at the nucleus and larger $R_N$.  The
new result, $|d(^{199}\mathrm{Hg})| \lsim 3.1 \times 10^{-29}$
e cm \cite{Hg}, provides the most stringent bounds on a variety of possible sources
of hadronic CPNC. 

There are plans for a new generation of experiments employing trapped stable or radioactive isotopes.
As traps allow more flexibility in the choice of nuclear and atomic spins, this
technique may open up opportunities to exploit
certain isotopes with enhanced polarizabilities
(though higher spins may also increase sensitivity to field inhomogeneities). 
For example, the CPNC mixing of the  160 eV 5/2$^-$-5/2$^+$ ground-state parity doublet in $^{229}$Pa is expected to enhance the
nuclear edm by a factor of $\sim 10^3-10^4$ \cite{Henley}.   Theoretical studies of the
collective enhancements of dipole moments of octupole-deformed nuclei, where parity doublets arise
naturally \cite{Doba}, helped motivate Argonne and KVI
proposals for $^{225}$Ra.  

\section{Precise Measurements of Weak Decays}
As approximately a dozen contributions to this conference discuss the use of
precise decay measurements to probe the weak interaction, I wish
there were more time available to discuss this field.  There are several motivations
for these difficult experiments:
\begin{itemize}
\item probing general properties of weak rates, such as universality, 
mixing angles (e.g., the extraction of $V_{ud}$ from Fermi $\beta$ decay or
from the neutron), and coupling strengths (e.g., the
pseudoscalar coupling $F_P$ or the second-class tensor coupling $F_T$);
\item constraining symmetry-breaking new interactions by their effects on muon
or neutron decay, such as the exotic P-even, pseudo-T-odd neutron-decay
$D$ coefficient
\begin{eqnarray}
 {d \omega \over dE_e d \Omega_e d \Omega_\nu} \propto p_e E_e (E_0-E_e)^2
&&\left[1 + a\vec{\beta}_e \cdot \hat{p}_\nu + A \vec{\sigma}_n \cdot \vec{\beta}_e \right. \nonumber \\
&&\left. + B \vec{\sigma}_n \cdot \hat{p}_\nu + b {m_e \over E_e} + D \vec{\sigma}_n \cdot 
(\vec{\beta}_e \times \hat{p}_\nu) \right] \mathrm{;~and}
\end{eqnarray}
\item verifying SM relations among decay parameters, e.g.,
\begin{equation}
a = {g_V^2-g_A^2 \over g_V^2 -3 g_A^2}~~~~~~~~~~A=-2 {g_V^2+g_Ag_V \over g_V^2-2g_A^2}~~~~~~~~~~B=2 {g_V^2-g_A g_V \over g_V^2-3 g_A^2}.
\end{equation}
\end{itemize}

Nuclei can be useful in such tests: selection rules can simplify constraints.  For example,
in a high-Q $0^+ \rightarrow 0^+$ $\beta$ decay, the 
back-to-back emission of the $e^+$ and $\nu_e$ is forbidden for a $V-A$ interaction
because of unbalanced angular momentum associated with the handedness of the
leptons.  The addition of scalar ($S,S^\prime$) interactions
\begin{equation}
H_\beta = \bar{\psi}_n \gamma_\mu \psi_p~\bar{\psi}_\nu C_V \gamma_\mu (1-\gamma_5) \psi_e~
+ ~\bar{\psi}_n \psi_p~\bar{\psi}_\nu (C_S + C_{S^\prime} \gamma_5) \psi_e
\end{equation}
produces leptons with identical chiralities, so that emission in the same direction is
forbidden, and back-to-back leptons preferred.   Thus the daughter nucleus recoil
momentum distribution (the observable) can be a sensitive test for nonzero
$C_S$ and $C_{S^\prime}$.  Among the interesting cases are $^{32}$Ar \cite{Ar} (where 
improved momentum resolution was achieved by measuring final-state delayed
protons) and $^{38}$K$^m$ \cite{K} (where
a magneto-optical trap allowed the low-energy recoiling nucleus to freely
escape to a detector). 

\section{Flavor and Total Lepton Number}
The discovery of neutrino
oscillations (e.g., $\nu_e \rightarrow \nu_\mu$) demonstrates that flavor
is not conserved among the leptons and motivates further tests of
lepton flavor number and total lepton number nonconservation,
\begin{equation}
\sum_{in} l_e \neq \sum_{out} l_e~~~~~~~~~\mathrm{or}~~~~~~~~~\sum_{in} l_e+l_\mu+l_\tau \neq
\sum_{out} l_e+l_\mu+l_\tau.
\end{equation}
Total lepton number plays a
special role in descriptions of neutrino mass.

\begin{figure}
\centering
\includegraphics[width=13.5cm]{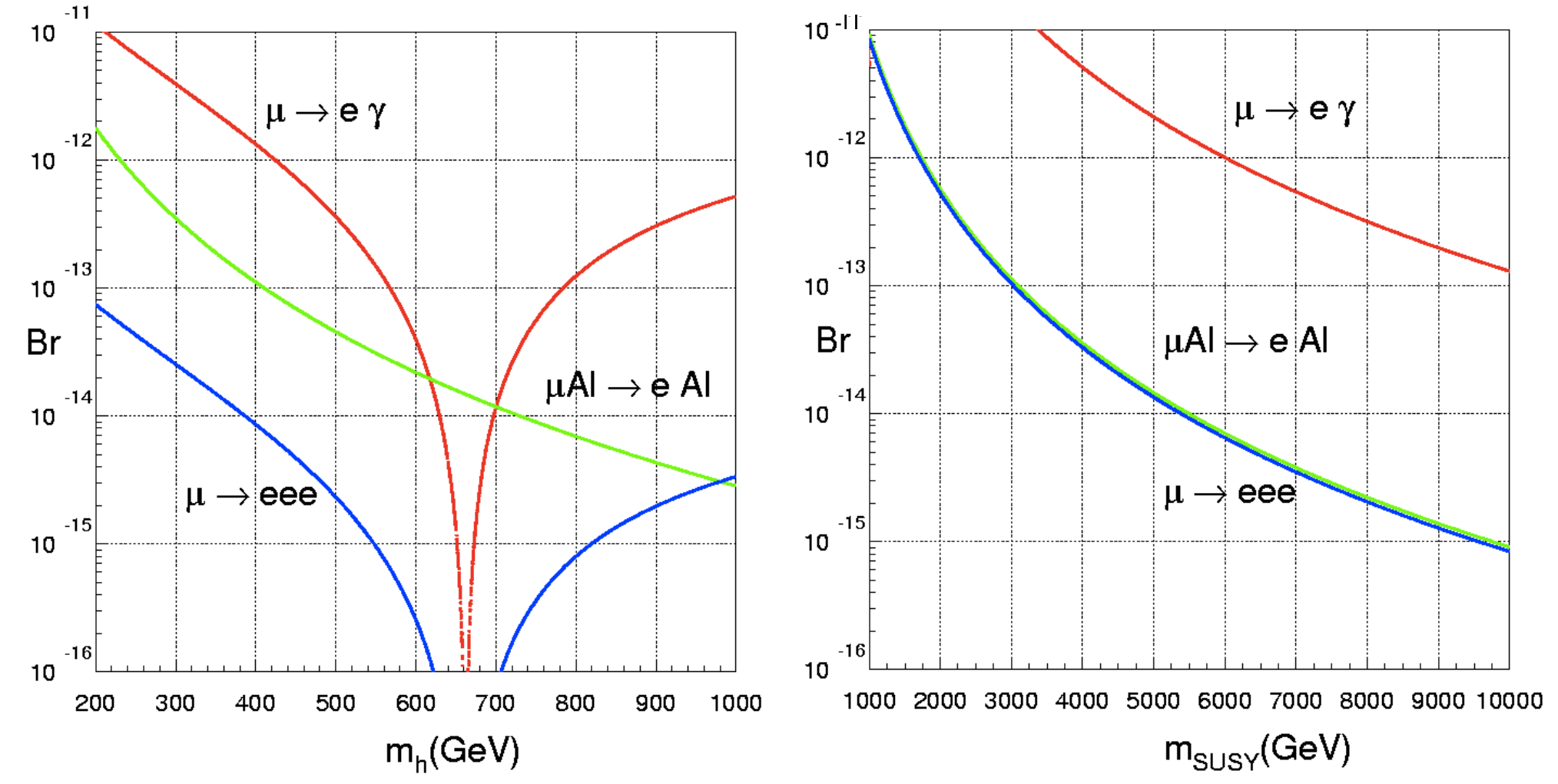}
\caption{Illustrations from \cite{CERN,Paradisi} of the expected branching ratios for Higgs-mediated 
and gaugino-mediated LFV, vs. the respective Higgs boson and SUSY masses.}
\label{fig:muon}
\end{figure}

As the sources of lepton flavor violation (LFV) are varied, the relative sensitivities to new physics
are highly model dependent.  Classic low-energy tests include $\mu \rightarrow e + \gamma$,
$\mu + (N,Z) \rightarrow e + (N,Z)$, and $\mu \rightarrow e+e+e$, as well as
corresponding $\tau$ decays. Fig. \ref{fig:muon}
shows representative sensitivities for two LFV mechanisms.

Table \ref{tab:muon} lists some of the LFV branching ratio bounds that have been obtained
in the past two decades.  The generally favored $\mu^+ \rightarrow e^+ \gamma$ mode can be
mimicked by final-state accidentals, limiting the advantages of high-intensity muon beams.
In contrast, as beam intensity is often a limiting factor in searches for $\mu^- \rightarrow e^-$
conversion in nuclei, improved beams can lead to large increases in sensitivity.   
The conversion process requires good energy resolution,
$\lsim$ 1 MeV, to exclude the background from ordinary $\mu$ decay.

\begin{table}
\caption{Limits on LFV from experiments with muons \cite{CERN}}
\begin{tabular}{|l|c|c|c|}
\hline
~Mode~ & ~Bound (90\% c.l.)~ & ~Year~ & ~Experiment/Laboratory~ \\
\hline
~$\mu^+ \rightarrow e^+ \gamma$ & ~1.2$ \times 10^{-11}$~~&~2002~& ~MEGA/LAMPF~\\
~$\mu^+ \rightarrow e^+ e^+ e^- $& ~1.0$ \times 10^{-12}$~~&~1988~& ~SINDRUM I/PSI~\\
~$\mu^+ e^- \leftrightarrow \mu^- e^+$  & ~8.3$ \times 10^{-11}$~~&~1999~&~PSI~\\
~$\mu^-$Ti $\leftrightarrow e^-$ Ti~~& ~6.1$ \times 10^{-13}$~~&~1998~&~SINDRUM II/PSI~\\
~$\mu^-$Ti $\leftrightarrow e^+$ Ca$^*$~~& ~3.6$ \times 10^{-11}$~~&~1998~&~SINDRUM II/PSI~\\
~$\mu^-$Pb $\leftrightarrow e^-$ Pb~~& ~4.6$ \times 10^{-11}$~~ &~1996~&~SINDRUM II/PSI~\\
~$\mu^-$Au $\leftrightarrow e^-$ Au~~& ~7.0$ \times 10^{-13}$~~&~2006~&~SINDRUM II/PSI~\\
\hline
\end{tabular}
\label{tab:muon}
\end{table}

Both J-PARC and FermiLab have plans for next generation $\mu \rightarrow e$ conversion
experiments that will substantially improve limits on LFV.  These experiments will use pulsed
proton beams to remove pion backgrounds by timing, large-acceptance capture
solenoids to increase the $\mu$ flux,  and bent solenoids to transport the muons, removing
neutrals and separating charge.  The FermiLab experiment will use 8 GeV protons
from a new driver and has a branching ratio goal of $4 \times 10^{-17}$, while the
J-PARC experiment will use a 40 GeV proton beam and has a goal of $5 \times 10^{-19}$ \cite{CERN}.
This program will push LFV sensitivities for scalar exchanges from the
current level of $\sim$ 1 TeV to $\sim$ 10 TeV. 

\begin{figure}
\centering
\includegraphics[width=10cm]{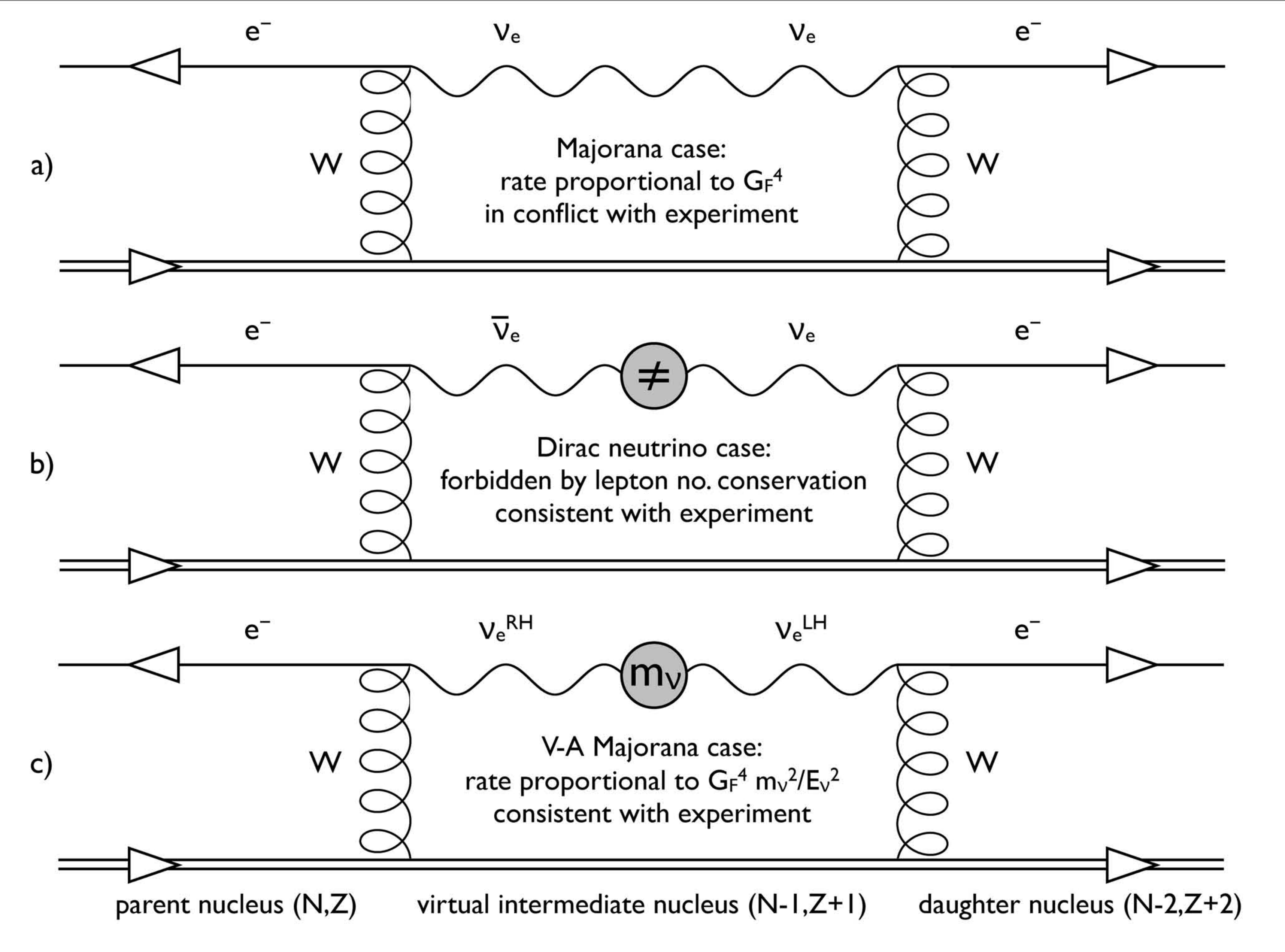}
\caption{Neutrinoless $\beta \beta$ decay scenarios:  a) the pre-1957 Majorana case,
which appeared to conflict with experiment; b) the Dirac case, where the process is
forbidden by lepton number conservation; and c) the Majorana case, where the
handedness mismatch is not total, due to the neutrino mass.}
\label{fig:bb}
\end{figure}

Tests of total lepton number are important to the description of massive neutrinos.
Neutrinos are unique among SM fermions in lacking an obvious charge or other additively conserved
quantum number that would reverse sign under particle-antiparticle conjugation.
Consequently, particle and antiparticle could be identical, $\nu=\bar{\nu}$.  That is,
the neutrino might be a Majorana particle, rather than Dirac ($\nu \perp \bar{\nu}$).  Prior to
1957 the absence of neutrinoless $\beta \beta$ decay appeared
to rule out a Majorana neutrino, since the process illustrated in Fig. \ref{fig:bb}a) would
lead to relatively rapid decay.  This result seemed to require a ``charge" -- lepton number --
to distinguishing $\nu$ and $\bar{\nu}$, and the assumption of conservation of that
charge to account for the absence of $\beta \beta$ decay,
\begin{equation}
l_e(e^-) = l_e(\nu_e)=+1~~~~~l_e(e^+)=l_e(\bar(\nu)_e)=-1~~~~~~~~~~\sum_{in} l_e = \sum_{out} l_e.
\end{equation}
The process in Fig. \ref{fig:bb}b) would then be forbidden, as there are no leptons in the
initial state, while the final state carries a net lepton number of two. 

But this conclusion ignores neutrino helicity:
a massless right-handed neutrino has the wrong handedness to be reabsorbed in the second
$\beta$ decay of Fig. \ref{fig:bb}a).  That is, absence of neutrinoless $\beta \beta$
tells us nothing about the Dirac or Majorana nature of the neutrino if the process is
independently forbidden by maximal PNC.

The discovery of neutrino mass, however, changes this argument.  Handedness is no
longer exact and thus does not forbid $\beta \beta$ decay.  Instead, the rate is suppressed
by a factor $(m_\nu/E_\nu)^2$, where
$E_\nu$ is the typical energy of the exchanged neutrino, as illustrated in
Fig. \ref{fig:bb}c).  If one can overcome this suppression factor by doing a very
sensitive experiment, $\beta \beta$ decay would be observed, provided Majorana
neutrinos exist.  Furthermore, the rate might tell us something about
the scale of neutrino mass -- important because
oscillation experiments only probe $m^2$ differences.

One might expect the neutrino to have both Dirac $\bar{\Psi}_R M_D \Psi_L$ and
Majorana $\bar{\Psi}_R^c M_R \Psi_R$ components.  The latter breaks the global
gauge invariance $\Psi \rightarrow e^{i \alpha} \Psi$ associated with a conserved
lepton number $l_e$, and thus can contribute to manifestly $l_e$-forbidden
processes like neutrinoless $\beta \beta$ decay.  As was discussed in Concha 
Gonzalez-Garcia's talk, there is a prejudice for such masses because they
also explain, via the seesaw mechanism, why neutrinos are light.  In the seesaw
mechanism the diagonalization of the mass matrix
\begin{equation}
\left( \begin{array}{cc} 0 & M_D \\ M_D & M_R \end{array} \right) ~\longrightarrow~
m_\nu^{\mathrm{light}} ~\sim~ M_D \left( {M_D \over M_R} \right)
\end{equation}
provides the needed ``small parameter" $M_D/M_R$ that explains why neutrinos are
so much lighter than other SM fermions (which can only have Dirac masses).
Thus light neutrinos are a reflection of the greater freedom available in building neutrino masses.  Current oscillation results
suggest $M_R \sim 3 \times 10^{15}$ GeV, a value near the GUT scale.

The possibility of discovering total-lepton-number violation is high because
\begin{itemize}
\item Nature likely makes use of Majorana masses;
\item atmospheric neutrino experiments suggest that the mass scale is not
unreasonably small, $m_\nu \gsim 0.05$ eV; and
\item extraordinary efforts are underway to mount massive new $\beta \beta$
decay experiments that will extend current sensitivities by an additional
factor $\sim$ 1000.
\end{itemize}

\section{Summary}
Precise tests of symmetries and conservation
laws, using low-energy techniques, remain one of our best windows on physics of
and beyond the SM, complementing the experiments performed at the energy
frontier.  Their utility derives from
\begin{itemize}
\item the many opportunities to isolate interesting interactions in both elementary
and composite systems, using angular momentum, parity, and kinematics;
\item exquisite experimental sensitivities (atomic shifts of $\sim 10^{-26}$ eV, $\beta \beta$
decay lifetimes of $10^{26}$ years);
\item level degeneracies and collective responses enhancing
interesting interactions in atoms and nuclei;
\item unique sensitivities, such as the GUT-scale reach of $\beta \beta$ decay;
\item the improving intensities of muon and cold neutron beams; and
\item the capacity of theory to connect what is learned at low-energies to both
astrophysics (e.g., neutrino mass) and accelerator physics (e.g., supersymmetry
at the LHC).
\end{itemize}

\section{Acknowledgment}
This work was supported in part by the Office of Nuclear Physics, US Department
of Energy.  I thank the organizers of PANIC08 for a most enjoyable meeting.

%
%
%

%

\begin{thebibliography}{00}

 \bibitem{JILA} C. S. Wood et al., \emph{Science} \textbf{275}, 1759 (1997); S. C. Bennett and 
C. E. Wieman, \emph{Phys. Rev. Lett.} \textbf{82}, 2484 (1999) 

 \bibitem{Derevianko} S. G. Porsev, K. Beloy, and A. Derevianko, arXiv: 0902.03351
 
 \bibitem{Young} R. D. Young, R. D. Carlini, A. W. Thomas, and J. Roche, \emph{Phys. Rev. Lett.}
 \textbf{99}, 122003 (2007)
 
 \bibitem{Erler} J. Erler and M. J. Musolf, \emph{Phys. Rev. D} \textbf{72}, 073003 (2005)
 
 \bibitem{DDH} B. Desplanques, J. F. Donoghue, and B. R. Holstein, \emph{Annals Phys.}
 \textbf{124}, 449 (1980)
 
 \bibitem{Dan} D. R. Phillips, M. R. Schindler, and R. P. Springer, arXiv:0812:2073;
 S. L. Zhu {\it et al.}, \emph{Nucl. Phys. A} \textbf{748}, 435 (2005)
 
 \bibitem{Adelberger} E. G. Adelberger and W. C. Haxton, \emph{Ann. Rev. Nucl. Part. Sci} \textbf{35}, 501 (1985)
 
 \bibitem{CERN} M. Raidal {et al.}, arXiv:0801:1826
 
 \bibitem{Hg} W. C. Griffith, M. D. Swallows, T. H. Loftus, M. V. Romalis, B. R. Heckel, and
 E. N. Fortson, arXiv:0901.2328
 
 \bibitem{Henley} W. C. Haxton and E. M. Henley, \emph{Phys. Rev. Lett.} \textbf{51}, 1937 (1983);
 O. P. Sushkov, V. V. Flambaum, and I. B. Khriplovich, \emph{Zh. Exp. Teor. Fiz.} \textbf{87},
 1521 (1984) 
 
 \bibitem{Doba} N. Auerbach, V. V. Flambaum, and V. Spevak, \emph{Phys. Rev. Lett.}
 \textbf{76}, 4316 (1996); J. Dobaczewski and J. Engel, \emph{Phys. Rev. Lett.} \textbf{94}, 232502 (2005)
 
 \bibitem{Ar} E. G. Adelberger {\it et al.}, \emph{Phys. Rev. Lett.} \textbf{83}, 1299 (1999)
 
 \bibitem{K} A. Gorelov {\it et al.}, \emph{Phys. Rev. Lett.} \textbf{94}, 142501 (2005)
 
 \bibitem{Paradisi} P. Paradisi, \emph{JHEP} \textbf{0608}, 047 (2006)
 

%
%
%
%
%
\end{thebibliography}
\end{document}